\documentclass[aps,prb,twocolumn,superscriptaddress,showpacs]{revtex4}
\usepackage{times}
\usepackage{amssymb,amsmath}
\usepackage{graphicx}
\usepackage{natbib}
\usepackage{multirow}
\usepackage{color}
\usepackage[a4paper,left=2cm,top=3cm,right=1cm,bottom=2cm]{geometry}

\begin{document}

\title{A highly accurate determination of absorbed power during nanomagnetic hyperthermia}

\author{I. Gresits}
\affiliation{Department of Non-Ionizing Radiation, National Public Health Institute, Budapest}
\affiliation{Department of Physics, Budapest University of Technology and Economics and MTA-BME Lend\"{u}let Spintronics Research Group (PROSPIN), Po. Box 91, H-1521 Budapest, Hungary}

\author{Gy. Thur\'{o}czy}
\affiliation{Department of Non-Ionizing Radiation, National Public Health Institute, Budapest}

\author{O. S\'agi}
\affiliation{Department of Physics, Budapest University of Technology and Economics and MTA-BME Lend\"{u}let Spintronics Research Group (PROSPIN), Po. Box 91, H-1521 Budapest, Hungary}

\author{I. Homolya}
\affiliation{Department of Physics, Budapest University of Technology and Economics and MTA-BME Lend\"{u}let Spintronics Research Group (PROSPIN), Po. Box 91, H-1521 Budapest, Hungary}

\author{G. Bagam\'{e}ry}
\affiliation{Mediso Medical Imaging Systems Ltd., Budapest, Hungary}

\author{D. Gaj\'{a}ri}
\affiliation{Mediso Medical Imaging Systems Ltd., Budapest, Hungary}

\author{M. Babos}
\affiliation{Mediso Medical Imaging Systems Ltd., Budapest, Hungary}

\author{P. Major}
\affiliation{Mediso Medical Imaging Systems Ltd., Budapest, Hungary}
 
\author{B. G. M\'arkus}
\affiliation{Department of Physics, Budapest University of Technology and Economics and MTA-BME Lend\"{u}let Spintronics Research Group (PROSPIN), Po. Box 91, H-1521 Budapest, Hungary}

\author{F. Simon\email{f.simon@eik.bme.hu}}
\affiliation{Department of Physics, Budapest University of Technology and Economics and MTA-BME Lend\"{u}let Spintronics Research Group (PROSPIN), Po. Box 91, H-1521 Budapest, Hungary}

\begin{abstract}
Absorbed power of nanoparticles during magnetic hyperthermia can be well determined from changes in the quality factor ($Q$ factor) of a resonator, in which the radiofrequency (RF) absorbent is placed. We present an order of magnitude improvement in the $Q$ factor measurement accuracy over conventional methods by studying the switch-on and off transient signals of the resonators. A nuclear magnetic resonance (NMR) console is ideally suited to acquire the transient signals and it also allows to employ the so-called pulse phase-cycling to remove transient artifacts. The improved determination of the absorbed power is demonstrated on various resonators in the 1-30 MHz range including standard solenoids and also a birdcage resonator. This leads to the possibility to detect minute amounts of ferrite nanoparticles which are embedded in the body and also the amount of the absorbed power. We demonstrate this capability on a phantom study, where the exact location of an embedded ferrite is clearly detected.
\end{abstract}

\maketitle

\section*{Introduction}
Nanomagnetic hyperthermia, NMH, \cite{{PankhurstReview2003},{PankhurstHyperthermia},{KUMAR2011789},{ANGELAKERIS20171642},{GIUSTINI2010},{KRISHNAN2010},{BEIK2016205}} emerged as a potential tool for tumor treatment in cancer therapy. It involves a targeted delivery of ferrite nanoparticles to the affected tissues and its heating with an external RF magnetic field which warms selectively and efficiently the embedding tissue only. The success of NMH relies heavily on several key medical factors\cite{{KUMAR2011789},{GIUSTINI2010},{ANGELAKERIS20171642},{PERIGO20150204}} such as the affinity of the tumor tissue to overtemperature and how specifically the ferrite is delivered to the desired location. On the physics side, the method depends on the accurate control and knowledge of the power which is dissipated by the ferrite. To obtain this information, most methods involve modeling of the exciting RF magnetic field and this information is combined with the knowledge of the magnetic properties of the delivered ferrite \cite{{GARAIO2014432},{GARAIO20142511},{0957-4484-26-1-015704},{CONNORD2014},{CARREY2011}} or the delivered heat is determined from calorimetry \cite{{GARAIO2014432},{0957-4484-26-1-015704},{WANG2013},{ESPINOSA20161002},{Bae2012}}, which however is an invasive and inaccurate method. 

We recently reported\cite{GresitsSciRep} a method to obtain \emph{directly} the power dissipated in the ferrite without any prior knowledge about the RF magnetic field strength or the ferrite properties. The method is based on the measurement of the quality factor, $Q$, of an RF resonator with and without the ferrite sample. The accuracy of the method for the dissipated power relies on the accurate determination of $Q$. A highly sensitive measurement could lead to e.g. the localization of minute amounts of ferrite and to study its diffusion under \textit{in vivo} conditions, a better assessment of specific absorbed power in the NMH materials or to study non-linear absorption effects in the ferrites. 

Conventional measurement of a resonator $Q$ factor is performed by sweeping the excitation frequency and studying the reflected signal \cite{LuitenReview,KajfezReview,KajfezVNA}. Albeit readily implemented, the frequency swept method is known to have a low accuracy for $Q$ and it is limited in measurement time to a few 100 ms thus this method does not enable the study of dynamic absorption effects (Refs. \onlinecite{PetersanAnlage,LuitenReview,KajfezReview}). Rather than measuring in the frequency domain, resonator parameters can be determined in time-domain measurements \cite{Gallagher,KomachiTanaka,Amato,EatonTransient}; then the resonator is excited with a pulsed carrier signal, whose frequency is close to the resonator eigen-frequency, $f_0$. Importantly, during both the switch-on and off, the resonator oscillates at its eigen-frequency, $f_0$, and the transient decay envelope has a time constant of $\tau=Q/2\pi f_0$. The transient signal can be Fourier transformed following a superheterodyne detection, which yields directly the resonance profile. This scheme is widely used in the study of high-$Q$ optical resonators \cite{optics1,optics2,optics3,optics4} and it was also implemented to measure the properties of microwave resonators \cite{GyureRSI2015,GyureRSI2018}.

In general, the time-domain measurements (like Fourier-transform NMR\cite{Ernst} and FT-IR spectroscopy) have two advantages: improved accuracy (or Connes advantage \cite{Connes}) since the measurement is traced back to a stable clock-frequency and simultaneous measurement (the Fellgett or multiplex advantage \cite{Fellgett}) of the resonance curve is attained. In fact, the resonator transient (also known as resonator ring-down) is usually an unwanted side effect and a well-known hindrance in low-frequency NMR \cite{HoultRSI}. It results in a "dead-time" in pulsed magnetic resonance and various schemes have been devised to reduce it\cite{HoultRSI, FukushimaRoeder}. However, the very same instrumentation of a standard pulsed NMR instrument, known as an NMR console, allows a direct measurement of the resonator transients. 

The present study is motivated by the quest for improved $Q$ determination accuracy with the goal to improve the absorbed power measurement during hyperthermia. We present that studying transients for RF resonators with $f_0=1-30\,\text{MHz}$ (including simple solenoid and MRI birdcage coils) does result in an order of magnitude more accurate $Q$ determination than the conventional frequency swept methods. We show that a commercial NMR console is readily adapted for this study without any modifications. Its use even allows to employ the so-called phase-cycled pulse schemes, which leads to the elimination of some transient artifacts. We demonstrate the low level of noise in actual measurements and also the performance of the method on a phantom study, which allows to locate a small amount of ferrite.

\section*{The instrument setup and its performance}

\begin{figure}[htp]
\begin{center}
\includegraphics[scale=.45]{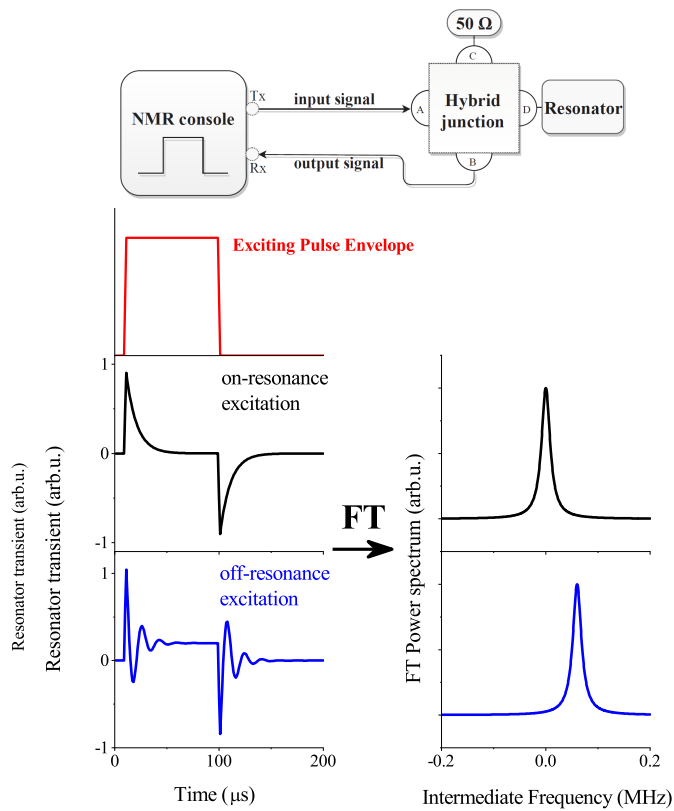}
\caption{The setup used to determine the resonator $Q$ and $f_0$ from the switch on/off transients (upper panel). The NMR console outputs the exciting pulse sequences and also detects them. A hybrid junction acts as duplexer to separate the excitation and the signal reflected from the resonator which is referenced with respect to 50 $\Omega$. The transient scheme is also shown: for both switch on and off, a transient signal is observed (lower panel, left). It is a single exponential when $f=f_0$ but it oscillates when $f\neq f_0$. the corresponding Fourier transform signals are also shown as a function of the intermediate frequency $f-f_0$ (lower panel right).}
\label{Fig:Setup}
\end{center}
\end{figure}

Fig. \ref{Fig:Setup}. shows the block diagram of the instrument used to measure the resonator quality factor in the time domain. A commercial NMR console (obtained from Mediso Medical Imaging Systems Ltd.) creates the pulses which excite the resonator and it also detects the reflection from the resonator in the time domain. A hybrid tee junction (M/A-COM HH-108 for 0.2-35 MHz and M/A-COM HH-107 for 2-200 MHz) separates the exciting and reflected waves from the RF resonator. The NMR console has a digitizing bandwidth of 4 MHz, thus it can acquire transients up to this bandwidth. A conventional reflectometry setup, which involves a swept frequency source and a power detector, which rectifies the RF signal, was used for comparison. Fig. \ref{Fig:Setup}. lower panel also depicts the transient detection scheme: the resonator is excited with a pulsed carrier signal whose frequency, $f$, is close to the resonator eigen-frequency, $f_0$. Typical pulses are $100~\mu\text{s}$ long for the irradiation (to achieve steady-state RF excitation in the resonator), followed by another $100~\mu\text{s}$ long acquisition, when the switch-off transient is detected. These pulse sequences are repeated after a delay of a few ms.

For both switch on and off, the resonators reflects a transient signal which differ only in their phase \cite{EatonTransient,GyureRSI2015} and has a frequency $f_0$ and decay with $\tau$. Note that the NMR console detects in (or $\mathrm{Re}$) and out-of-phase (or $\mathrm{Im}$) signal (also known as quadrature detection) but only one component is shown in the figure to retain clarity. The meaningful spectral data is obtained from the power spectrum of the FT data, i.e. $\mathrm{Re}^2+\mathrm{Im}^2$ and is shown in the figure.

In the superheterodyne detection scheme of the NMR console, the time-dependent signal is downconverted with $f$, it thus appears as an exponential function when $f\approx f_0$ or an oscillating function when $f\neq f_0$. When Fourier transformed, both the switch on and off transients yield a Lorentzian function with respect to the intermediate frequency (IF), $f-f_0$. 

\begin{figure}[htp]
\begin{center}
\includegraphics[scale=.40]{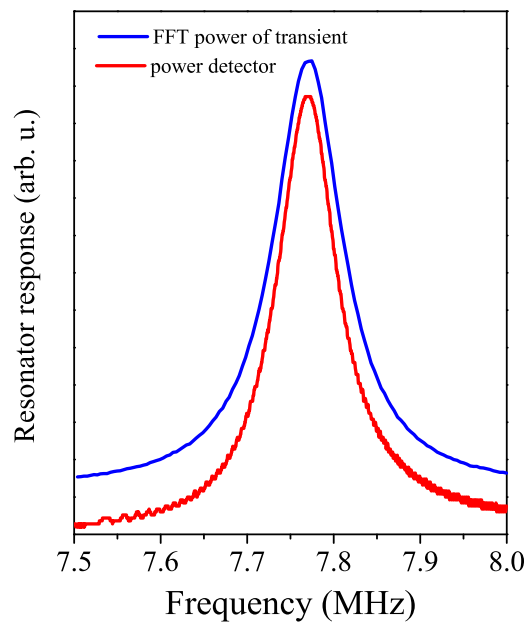}
\caption{Comparison of the $Q$-factor measurement using the conventional frequency swept power detector detection and the Fourier transformed transient signal. Note the larger noise for the power detector measurement, this includes the digitalization noise of the oscilloscope.}
\label{Fig:NMR_vs_detector}
\end{center}
\end{figure}

In Fig. \ref{Fig:NMR_vs_detector}., we show the time-dependent resonator transients as detected with the NMR console. The detection was performed in quadrature which allows a Fourier transformation of the signal, which yields directly the resonator curve around the intermediate frequency, IF. This is also shown on the lower panel in Fig. \ref{Fig:NMR_vs_detector}. The LO frequency is added to obtain the resonance curve on an absolute frequency scale. We performed measurements with a conventional, frequency swept method, in order to validate the present measurements. The two curves match well as demonstrated in Fig. \ref{Fig:NMR_vs_detector}. This means that the transient detection methods also yield the same kind of data such as the conventional measurement technique.

\begin{figure}[htp]
\begin{center}
\includegraphics[scale=.45]{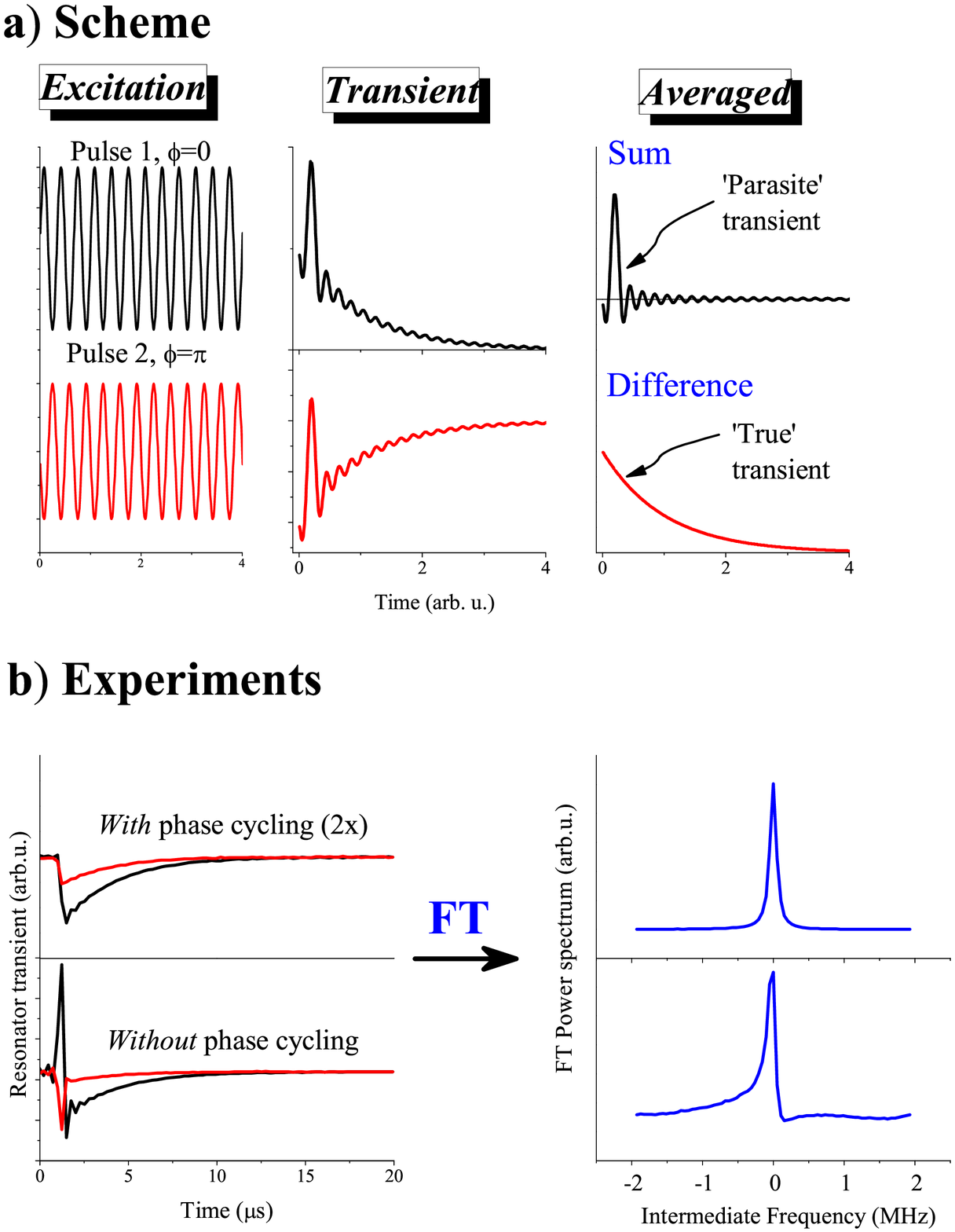}
\caption{a) The schematics of the phase cycling method as applied in the present measurement. Two consecutive excitations with opposite phase are applied to the resonator, and the resulting transient curves are subtracted to eliminate the unwanted parasitic transient effects (these are also known as \emph{ringing} in the NMR literature). b) Transients detected \textit{with} and \textit{without} the phase cycling method. The earlier signal is magnified by a factor 2. The corresponding Fourier transformed signals are also shown. Note the effect of the parasitic transients on the resulting resonator signal.}
\label{Fig:PhaseCycling}
\end{center}
\end{figure}

The use of the NMR console allows to implement the so-called phase cycling experiments, which is customary in NMR to get rid of the instrumental artifacts. Such artifacts include a DC offset of the digitizer or an imbalanced amplification in the two quadrature channels. As an example, the DC offset is tackled by exciting the NMR nuclei with pulses in the opposite RF phase and subtracting the signals after digitization. Imbalances in the quadrature detection are tackled by exciting the nuclei by cycling the RF phase by 90 degrees in consecutive excitation phases and cycling the phase of the digital downconversion by the same value. This scheme leads to the generic name, phase cycling, of the method. 

In our case, the dominant artifact is a peak followed by some parasitic "ringing" which appears when the pules are either switched on or off. This can be tackled by alternating the RF phase of the exciting pulse by 180 degrees: the parasitic signal is not sensitive to the RF phase, whereas the resonator transient is also rotated by 180 degrees as it is shown in Fig. \ref{Fig:PhaseCycling}a. When the resulting transients are subtracted accordingly only the desired transient is observed and the parasitic signal is eliminated. The effect of this phase cycling scheme is demonstrated in Fig. \ref{Fig:PhaseCycling}b.: the measured transient signal is free from any parasitic signal and its Fourier transform in a regular Lorentzian curve. In contrast, the unwanted signal appears without phase cycling and the corresponding Fourier transformed signal is also distorted.

It was established earlier \cite{GyureRSI2015,GyureRSI2018} that the appropriate errors of the $f_0$ and $Q$ determination accuracies are the following quantities:

\begin{align}
\delta\left(Q\right) := \frac{\sigma \left(Q\right)}{\overline{Q}}\,; \delta\left(f_0\right):=\frac{\sigma \left(f_0\right)}{\overline{\Delta f}},
\label{Eq:ErrorDefinition}
\end{align}
where $\overline{Q}$ and $\overline{\Delta f}$ are the mean values of $Q$ and the resonator bandwidth $\Delta f$, respectively. $Q=f_0/\Delta f$, where $f_0$ is the resonator frequency\cite{GyureRSI2015}. We note that the error of $f_0$ is not $\sigma(f_0)/f_0$ as it would be intuitive at first sight. This quantity would overestimate the accuracy for an ultra-high frequency resonator with a moderate $Q$ factor.

When comparing different measurement methods, a normalization with the measurement time is also important and we present data which is normalized to 1 second, thus the data is given in units of $1/\sqrt{\text{Hz}}$. We found that the conventional frequency swept method results in about $\delta\left(Q\right)\approx \delta\left(f_0\right)\approx 2\cdot 10^{-3} \cdot 1/\sqrt{\text{Hz}}$. In contrast, under the most optimal settings, the transient detection method results in about 20 times smaller $Q$ and $f_0$ determination errors, as small as $\delta\left(Q\right)\approx \delta\left(f_0\right)\approx 10^{-4} \cdot 1/\sqrt{\text{Hz}}$. The details of the most optimal transient acquisition settings, as well as a discussion of the error sources in terms of stochastic and drift-like errors, is presented in the Supplementary Materials.

\section*{Application to detect a minute amount of buried ferrite}

The ultra-high sensitivity of the present method opens the way for a number of applications in ferrite based hyperthermia of which we envisage a few. First, it allows to detect minute amounts of ferrite and the small amount of absorbed power in a realistic \emph{in vivo} animal model study. We previously calculated\cite{GresitsSciRep} that the smallest amount of detectable magnetite is 6 mg which is too much for hyperthermia therapy even in small laboratory animal models. However, as the sensitivity of our method is about 20 times better than our previous one, thus the lowest detectable ferrite amount is about 0.3-0.5 mg which is typically employed in mice animal model studies\cite{{ESPINOSA20161002},{Heidari2016},{Huang2013}}. 

The high sensitivity of the present method makes it suitable to detect non-linear (e.g.: saturation) effects or the change of the absorbed power due to the change of sample parameters e.g. the sample temperature. We demonstrate its utility to detect the position of a small amount of ferrite whose location is otherwise unknown. Locating the ferrite in hyperthermia is of great importance in medical applications; it could assess the success of drug delivery targeting and it could also allow to better focus the heating RF irradiation. 

\begin{figure}[htp]
\begin{center}
\includegraphics[scale=.45]{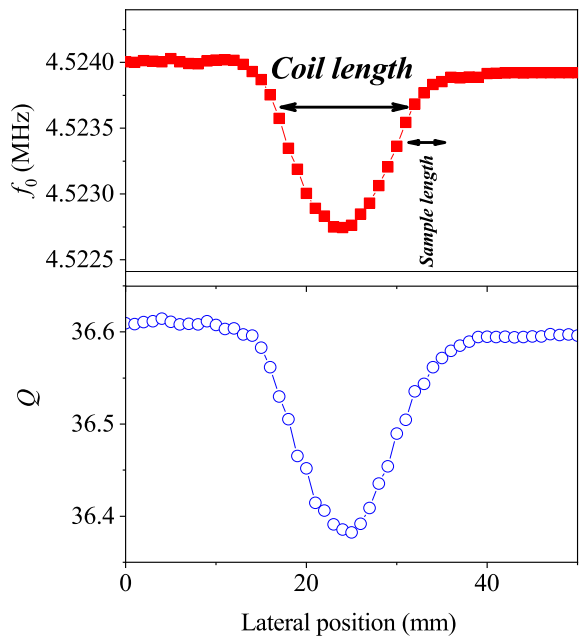}
\caption{Variation of $f_0$ and $Q$ when a small ferrite sample (containing about 1 mg of magnetite in solution) is moved across a solenoid. Each data point was recorded for 1 second. Arrows indicate the length of the solenoid and that of the sample. Note that both $f_0$ and $Q$ differ slightly on the two sides of the lateral movement due to an asymmetry in the sample holder.}
\label{Fig:Lateral_resolution}
\end{center}
\end{figure}

We show in Fig. \ref{Fig:Lateral_resolution}., the variation of $Q$ and $f_0$ when a small amount of magnetite (about 4 mm long, containing 1 mg of magnetite) is moved across the solenoid (with 14 mm length) of an RF circuit. We used 2 manual linear translation stages (Thorlabs GmbH) to achieve a lateral movement of 50 mm. Clearly, both circuit parameters are affected by the presence of the ferrite. Therefore locating an otherwise invisible ferrite with a good accuracy is made possible with the present technique.

\section*{Summary}
In summary, we presented a highly sensitive method to determine the power absorbed in nanomagnetic particles during radio-frequency irradiation. The method is based on monitoring the transient response of an impedance matched radio-frequency resonant circuit using a conventional nuclear magnetic resonance console. The latter technique allows the use of the so-called smart phase cycling schemes, which allows the artefact-free detection of the short transients. The method yields an unprecedented accuracy of the resonator quality factor and resonance frequency which reduces the amount of detectable ferrite during nanomagnetic hyperthermia. We demonstrate the utility of the method by sensitively detecting the location of a small amount of ferrite in a test tube.

\section*{Acknowledgements}
Support by the National Research, Development and Innovation Office of Hungary (NKFIH) Grant Nrs. K119442, 2017-1.2.1-NKP-2017-00001, and VKSZ-14-1-2015-0151 and by the BME Nanonotechnology FIKP grant of EMMI (BME FIKP-NAT), are acknowledged.


\appendix
\newpage
\pagebreak
\clearpage

\section{Noise of the measured parameters}

\begin{figure}[htp]
\begin{center}
\includegraphics[scale=.45]{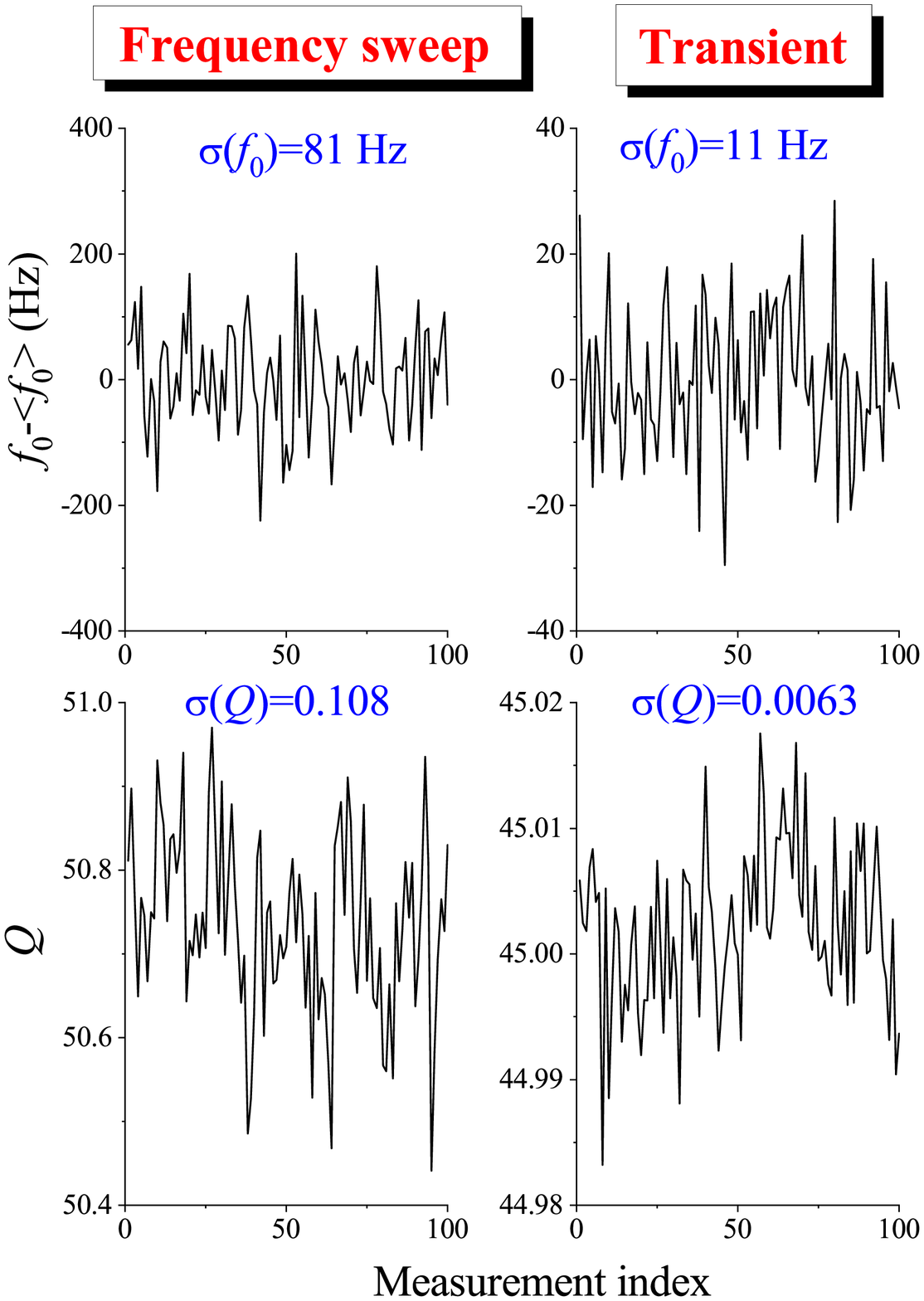}
\caption{Noise of the respective parameters as obtained with the conventional frequency sweep as well as the presented method for a measurement time of 1 sec/data point. A straight line was subtracted from the $f_0$ data due to a linear drift in both types of data and it is shown with respect to the mean values. The standard deviations, $\sigma\left( f_0\right)$ and $\sigma\left( Q\right)$, are given. We note that these values are about one sixth of the apparent peak-to-peak deviation from the respective mean values. Note the different scales for the two types of measurements.}
\label{Fig:Noise_properties}
\end{center}
\end{figure}


\begin{table*}[]
    \begin{tabular}{|c|c|c|c|c|c|c|}
    \hline
    \begin{tabular}[c]{@{}c@{}}Number of transient \\averages\end{tabular} & \begin{tabular}[c]{@{}c@{}}Measurement time\\ (seconds)\end{tabular} & \begin{tabular}[c]{@{}c@{}}Transient duration\\ ($\mu$s)\end{tabular} & $\delta Q$ & \begin{tabular}[c]{@{}c@{}}Stochastic $\delta$$Q$ to 1 s\\(in units of $1/\sqrt{\text{Hz}}$)\end{tabular} & \begin{tabular}[c]{@{}c@{}}Realistic $\delta$$Q$ to 1 s \\(in units of $1/\sqrt{\text{Hz}}$)\end{tabular} & \begin{tabular}[c]{@{}c@{}}Type of\\ measurement\end{tabular} \\ \hline
10 & 0.025 & 32.8 & $2.5\cdot10^{-4}$ & $4.5\cdot10^{-6}$ & $4.0\cdot10^{-5}$ & transient \\ \hline
100 & 0.22 & 32.8 & $1.6\cdot10^{-4}$ & $9.4\cdot10^{-6}$ & $7.7\cdot10^{-5}$ & transient \\ \hline
1000 & 2.17 & 32.8 & $4.6\cdot10^{-4}$ & $8.3\cdot10^{-5}$ & $6.8\cdot10^{-4}$ & transient \\ \hline
10 & 0.045 & 262.1 & $3.3\cdot10^{-4}$ & $1.7\cdot10^{-5}$ & $7.1\cdot10^{-5}$ & transient \\ \hline
100 & 0.41 & 262.1 & $1.5\cdot10^{-4}$ & $2.4\cdot10^{-5}$ & $9.5\cdot10^{-5}$ & transient \\ \hline
1000 & 4.21 & 262.1 & $3.5\cdot10^{-4}$ & $1.8\cdot10^{-4}$ & $7.2\cdot10^{-4}$ & transient \\ \hline
10 & 0.2 & 2097.2 & $3.2\cdot10^{-4}$ & $4.7\cdot10^{-5}$ & $1.4\cdot10^{-4}$ & transient \\ \hline
100 & 1.98 & 2097.2 & $2.5\cdot10^{-4}$ & $1.1\cdot10^{-4}$ & $3.5\cdot10^{-4}$ & transient \\ \hline
1000 & 1.967 & 2097.2 & $3.0\cdot10^{-4}$ & $4.2\cdot10^{-4}$ & $8.4\cdot10^{-4}$ & transient \\ \hline
128 & 7.37 & 1000 & $7.4\cdot10^{-4}$ & $2.7\cdot10^{-4}$ & $2.0\cdot10^{-3}$ & Freq. sweep \\ \hline
\end{tabular}
\caption{The statistics and the errors of the $Q$ determination for the transient based and the frequency sweep methods. The $Q$ value was typically 45. The $\delta Q$ quantity is determined according to the definition in the main text. This quantity can be normalized by the duration of the transient itself to yield the "Stochastic $\delta Q$ to 1 s". Alternatively, the $\delta Q$ can be normalized by the true measurement time, which yields a realistic estimate of the attainable accuracy of the $Q$ determination. E.g. the corresponding data in the first row is obtained by multiplying the raw $\delta Q$ is multiplied by $\sqrt{10 \cdot 32.8~\mu\text{s}}$ and $\sqrt{25~\text{ms}}$, respectively.} \label{tab:table1}
\end{table*}

In Table \ref{tab:table1}., we show the measurement parameters as well as the observed error of the $Q$ determination, $\delta Q$. 
We measured each data point in Table \ref{tab:table1}. 100 times, in order to obtain a statistically appropriate empirical estimate of the standard deviation for both $Q$ and $f_0$: $\sigma \left(Q\right)$ and $\sigma \left(f_0\right)$, which is used to obtain $\delta\left(Q\right)$ and $\delta\left(f_0\right)$ as explained in the main text.

We also give the error of the $Q$ determination normalized by two different methods: the first considers the presence of the stochastic noise only and is normalized by the time spent with the acquisition of the transients. The other normalization considers the total amount of time spent to measure each data point, it is therefore a realistic estimate of the error encountered in a real-life situation. We can draw several conclusions from the table: the value of $\delta Q$ is little affected by averaging the transients for 10-100- or 1000 times. This is the result of other than stochastic noise sources (probably drift effects). This means that it makes no sense to average the transients for larger than about 10 times. The error is also not improved for measuring the transients for longer times as the transient decays for a maximum of about $10-20~\mu\text{s}$ and in the rest of the measurement, only noise is acquired. The actual durations indicate that the spectrometer has a minimal acquisition time of about 2 ms for each averaging, which can not be further reduced. Given that our typical transient duration is $10-20~\mu\text{s}$, in principle a perfect instrument with zero acquisition- (or dead-) time would allow to improve $\delta Q$ by an additional factor of about 10.

\end{document}